# The illusory simplicity of the feedforward pass: evidence for the dynamical nature of stimulus encoding along the primate ventral stream

Daniel Anthes[1], Sushrut Thorat[1], Anna Mitola[2,3], Paolo Papale[2,3], Peter König[1], Tim C Kietzmann[1]

[1]Institute of Cognitive Science, University of Osnabrück, Osnabrück, Germany.
[2]Department of Medicine and Surgery, Neuroscience Unit, University of Parma, Parma, Italy.
[3]Department of Vision & Cognition, Netherlands Institute for Neuroscience (KNAW), Amsterdam, the Netherlands.

## Abstract

In studying primate vision, a large body of work focuses on the first feedforward sweep. During this initial time window, information is thought to pass through ventral stream regions in a stage-like fashion in an effort to extract high-level information from the retinal input. Consequently, electrophysiological analyses commonly focus on spatial response patterns, either by averaging data in time, or by applying decoders in a temporally local fashion. By analysing data recorded simultaneously across multiple arrays placed along the macaque ventral stream, we here show that this prior approach may be missing key aspects of information encoding. First, time-resolved, multivariate analyses of information transfer between V4 and IT reveal temporally and semantically varied information content as being exchanged within the first 100ms of processing. Second, by employing recurrent neural network (RNN) decoding techniques that extend across the temporal domain, we demonstrate that the neural pattern dynamics themselves carry categorical information far beyond the spatially encoded information available at any given time point. These findings challenge the prevailing view of a single, stage-like feedforward process and suggest that even the earliest parts of visual processing are better characterised as a spatiotemporally evolving process that encodes information in its dynamics rather than purely spatial response patterns.

## Introduction

When visual information arrives at the cerebral cortex, a sequence of computations is triggered across the ventral visual stream in service of extracting meaning from the retinal input. In this process, a central computational role is commonly attributed to the first, rapid cascade of information transfer. This process starts in early visual cortex (V1), and leads up to inferotemporal cortex (IT)[1], transforming 'low-level' representations into a more abstract, categorical code[2]. As a result of the assumption that this feedforward pass is a stage-like information transfer across regions, studies of the process often focus on spatial activity patterns, such as patterns obtained by taking the averages of activity over a temporal window of 100ms length, suitably placed after stimulus onset.

Here, we investigate this core assumption of a single feedforward transfer of information by analysing neural responses to a total of 1000 natural images, sampled from 100 natural object categories. Neural responses were recorded across multiple arrays positioned along the ventral stream of two macaques. Our time-resolved analyses of inter-area information flow between V4 and IT reveal multiple time points at which V4 activity is predictive of activity at IT recording sites. Remarkably, while our analyses are all performed within a time window commonly averaged in studies of the initial "feedforward" interval, early predictions are found to be qualitatively and quantitatively different from late predictions. Comparisons to representations in artificial neural network models demonstrate that models of early information transfer are characterized best by low-level information, whereas models of late information transfer are better characterized by more abstract, categorical representational geometries. Finally, decoding analyses by linear and recurrent probes call into question the conceptualisation of IT population activity as converging to a stable code that can then be read out from downstream areas. Instead, we find that the population dynamics themselves are highly informative of object category, enabling significant boosts in decoding performance. Our results jointly call into question the idea of a single feedforward information transfer that results in a stable population code in IT, but rather emphasise the importance of treating the ventral stream as a dynamical system, even during the earliest parts of information processing.

# Results

### Simultaneous array recordings across the ventral stream in response to a structured set of natural stimuli

Local field potentials (LFP) were simultaneously recorded from 16 (15 in a second individual) 64-channel Utah arrays, implanted across the ventral visual stream of two macaque monkeys. Eight (7) arrays were placed in V1, 3 (4) arrays in V4 and 5 (4) arrays in IT (**Fig. 1A**). Data were recorded during a passive viewing experiment in which the monkeys repeatedly viewed a subset from the THINGS dataset[3] consisting of 1000 natural images from 100 categories (**Fig. 1B**). Data for monkey F are shown in the main text. All results are replicated for monkey N in the supplemental materials (**Fig. S1-3**). LFPs were chosen for the main analyses due to their higher SNR compared to multi-unit activity in this dataset (**Fig. S4**).

### Staggered onset of stimulus-related information

Information processing across the ventral stream is commonly studied by analysing neural responses in each region separately. Applying this approach to our data, and in line with previous reports[2,4–6], we observe staggered onsets of reliable stimulus responses with increasing latency from V1 to V4, and on to IT. Similarly, peak category decoding performance shows increasing delays from V1 to IT. Best category decoding was observed in IT, peaking around 150ms post stimulus onset (**Fig. 1C-E**). These observations have historically provided the basis for the view of a stage-like, hierarchical feedforward information transfer across the ventral stream, with categorical information becoming available in IT with latencies compatible with the end of the feedforward sweep.

### At least two phases of V4 - IT communication during time-windows that are commonly considered a single feedforward sweep

Having investigated each region of interest independently, we turn to analyses of information transfer across regions in the ventral stream. We focus on the interactions between V4 and IT, which have been ascribed a central role in object categorisation. We characterise the feedforward interaction between V4 and IT with a set of linear models that predict neural activity measured at each channel in IT, from all channels in V4, taking into account a conduction delay. We fit models for a wide range of timepoints in the target array and various conduction delays from predictors in V4. For each time *t* and delay *d* two models are fit: The first model predicts activity in the target array at time *t* from activity in the same array at an earlier time (*t* - *d*; base model). The second model has an additional set of predictors from activity in V4 taken at the same time (*t* - *d*) (**Fig. 1F**). To identify timepoints and delays for which V4 adds predictive power beyond information that was already available in the target array, we report the difference in predictive performance between the two models. Contrary to the expectation of a single stage-like information transfer from V4 to IT, we observe two separate time windows in which V4 activity adds to the predictability of IT (**Fig. 1F**). Specifically, V4 activity recorded as early as 46ms predicts IT at approximately 74ms post stimulus onset. A second broader peak of V4 activity, recorded around 80ms, was observed to predict IT activity approximately 40 ms later. This phenomenon was observed across all arrays tested. The finding of two separate timepoints, 'early' and 'late', is noteworthy, as both appear during a time window commonly ascribed to the phase in which a single feedforward pass of information is performed in the system.

### Early and late feedforward predictions are driven by significantly different information

While the observation of multiple time points at which V4 is predictive of IT suggests different phases of information transfer, it does not necessarily imply different information content. To investigate this possibility, we chose one representative model-timing for each of the two communication time-windows (referred to as early and late above). The 'early models' predict arrays in IT 74ms from V4 46ms, and the 'late models' predict arrays in IT at 120ms from V4 at 80ms. We find that 'early' and 'late' models are qualitatively and quantitatively different in a number of ways. To compare the two phases, we first compared the underlying model weight matrices and their generalisation across time (cosine distance). In doing so, we focus on the parts of the weight matrix that correspond to the predictions originating from V4. For both the 'early' and 'late' reference models, we observe a temporal neighbourhood where dissimilarity between pairs of weight matrices is low, indicating similar mappings from V4 to arrays in IT in that temporal neighbourhood. However, distances to the weight matrices of the respective other reference model and their neighbourhood are large, indicating that, while weight matrices for 'early' and 'late' models form respective temporal clusters, they are dissimilar to each other (**Fig. 1G**).

To characterise the representational content transferred at the 'early' and 'late' time points, we performed a representational similarity analysis (RSA[7]). For this, we computed representational dissimilarity matrices (RDMs, 1000x1000 entries corresponding to the stimuli shown) based on the IT target array signals, as predicted by V4. We find that, across IT target arrays, all 'early' model predictions exhibit similar representational geometries. At the same time, all 'late' model predictions exhibit similar RDMs. However, the representational geometries differ significantly between early and late predictions, even in the same target array (**Fig. 1H**, $p = 0.0$, one-sided permutation test, 10.000 samples).

Finally, to assess the temporal development of the 'early' and 'late' phases, we used the respective linear models to predict the entire temporal trajectory of neural activity in IT target arrays. We then used these two V4-based predictions of IT responses to assess their unique contribution in explaining the real IT representational dynamics (RDM movies). Similar to Kietzmann et al.[8], we plot the unique variance explained across time, but here for the early and late model predictions, respectively. Our results indicate a representational handover in IT, as early and late models uniquely explain variance in IT array representational geometries at different points in time (**Fig. 1I,** cluster-based permutation test on the unique variance explained across arrays, comparing explained variance trajectories for predictions from 'early' models against 'late' models, 1024 permutations, one-sided F-statistic. Early cluster 56-96ms, p=0.005, late cluster 106-152ms, p=0.012). This suggests that the dynamics of IT can be explained by a changing readout of activity in V4.

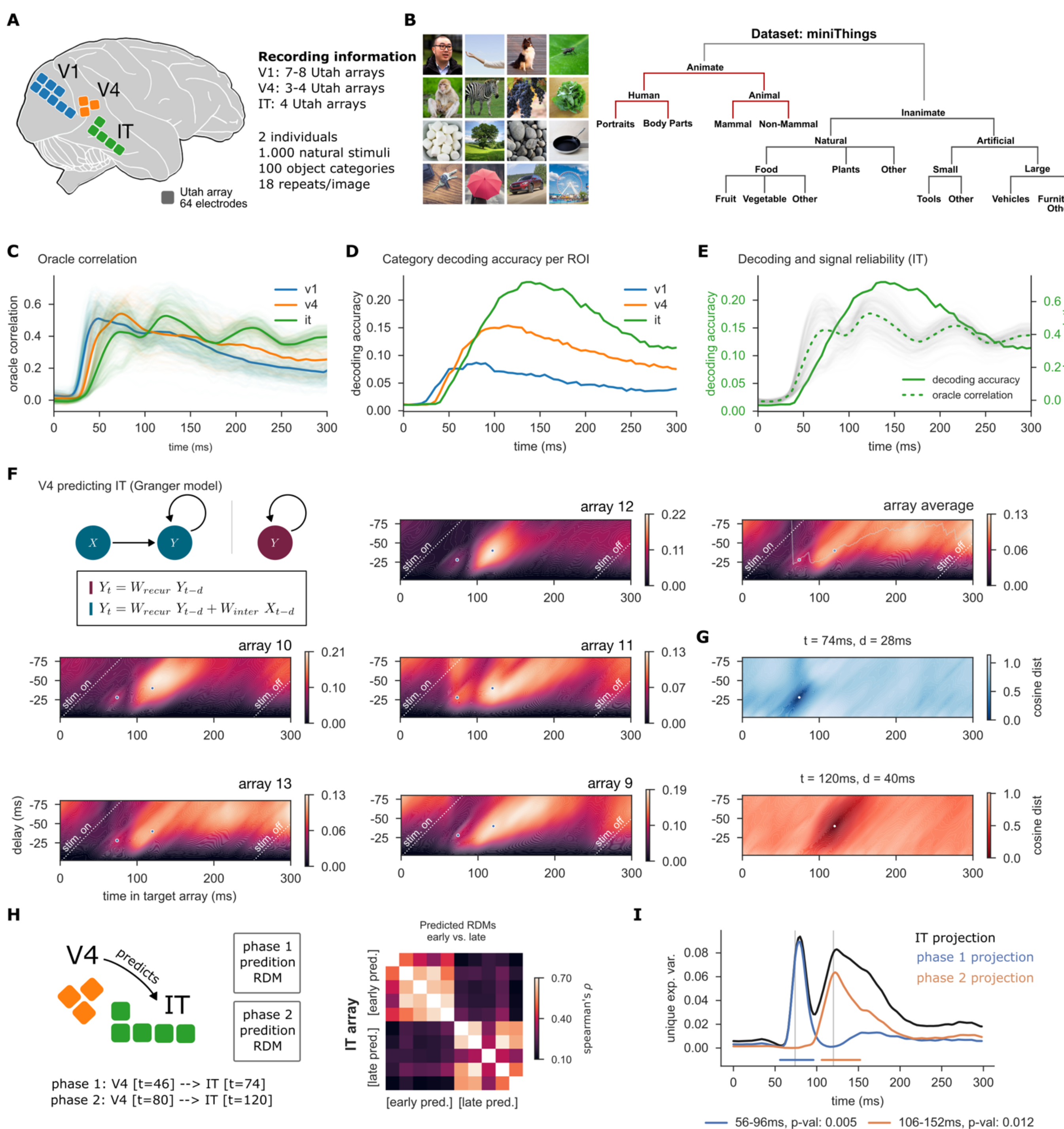

**Fig. 1**: **Predicting information transfer from V4 to IT during feedforward processing of visual information. A,** Two macaque monkeys were implanted with multiple electrode arrays across V1, V4, and IT. LFPs were recorded while monkeys passively viewed natural stimuli. **B,** Stimuli are a curated subset of THINGS[3] drawn from 100 categories with 10 exemplars each. Categories form a hierarchy of broad to specific 'super categories', sampling the diversity of THINGS. **C - E,** ROI-wise analyses resolved in time. The onset of stimulus-related information measured with oracle correlation, and categorical information measured with linear probes are both staggered in time with increasing latency from V1 to IT. **F,** Modelling inter-area information transfer. Neural activity at each of 5 arrays implanted in IT can be predicted from arrays in V4 at multiple times after stimulus onset. For each target array, there is an 'early' and 'late' cluster of timepoints where V4 predicts IT. One set of models per cluster is selected for further analysis: The 'early' models predicting IT at 74ms from V4 at 46ms are indicated with green points. The 'late' models predicting arrays in IT at 120ms from V4 at 80ms are indicated with blue points. **G,** Similarity of models over time. Surfaces show cosine distance between the weight matrices of the 'early' (top) and 'late' (bottom) models compared to models from all other times. There are separate clusters of similar models, 'early' and 'late', that are different from models in the other cluster. Distances are computed separately between models for each target array, and surfaces shown here are averages over arrays. **H,** representational similarity of 'early' and 'late' predictions of IT. Representational geometries of predicted signals in arrays in IT are similar for all models from the same time, but differ between 'early' and 'late' times. **I,** 'Early' and 'late' models uniquely predict representational geometry in IT in intervals around the times at which they were fit.

## Early and late feedforward predictions from V4 to IT align with lower- and higher-level ANN embeddings

To better understand the representational content of predictions from 'early' and 'late' models, we compare them to representations found across convolutional and fully connected layers of a ResNet18[9], trained on categorical classification of natural images. As previously, we derived representational dissimilarity matrices (RDMs) for each time point of the predicted activity at IT recording sites. This resulted in two RDM movies based on the early and late communication windows, respectively. We then compared each time point of both time series to the ANN model RDMs (**Fig. 2A**). We find that the predicted geometries from early models are best explained by ANN model RDMs derived from early ResNet layers, while predictions of the 'late' model are best explained by RDMs derived from late ResNet layers. These differences in representational content extend to the entire predicted time courses: early and late models make qualitatively different predictions favouring geometries better characterized by RDMs from early ResNet18 layers for predictions from the early model, and geometries better characterized by RDMs of later layers for predictions from the late model (**Fig. 2B**). This points towards two qualitatively different 'readouts' from V4 that predict 'low-level' and 'high-level' information in IT.

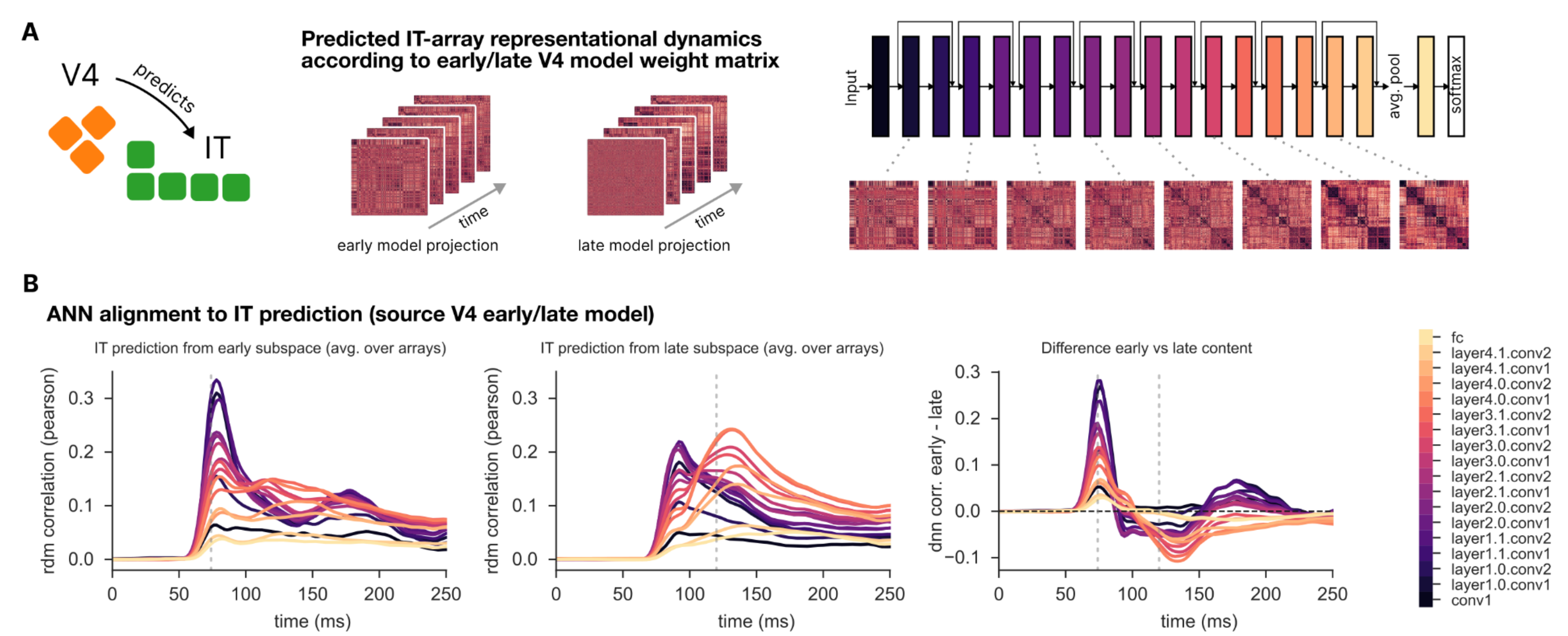

**Fig. 2**: **Characterizing representational geometry of early and late predictions of recording sites in IT with RSA. A,** Two sets of RDM movies for arrays in IT are predicted from V4 time courses using the 'early' and 'late' inter-area models selected for analysis (left). RDMs are extracted from multiple layers of ResNet18[9] as models for 'low-level' geometries (early layers) and 'categorical' geometries (late layers). **B,** Characterization of predicted representational geometries over time. RDM trajectories predicted by 'early' (left) and 'late' (middle) models of inter-area interaction are correlated with model RDMs extracted from ResNet18. Predicted time courses from 'early' models correlate most highly with early ResNet layers, while time courses predicted from 'late' models show increased correlation with later layers. Dashed vertical lines indicate times for which models were fit. Taking the difference between the two panels (right) indicates that representational geometries predicted from 'early' and 'late' models are qualitatively different for late times as well, suggesting that differences in population activity falling onto either readout from V4 persist for the duration of stimulus presentation. Trajectories of population activity measured from IT directly have been shown to follow a similar progression from low- to high-level geometries[10]. An analysis of the representational content of population activity recorded from IT is presented in Fig. S5 for both LFP and MUA.

### Temporal dynamics in IT carry categorical information

Our results so far suggest that what is commonly described as a single ‘feedforward’ information transfer from V4 to IT may more accurately be characterized as a dynamical, time-varying process. This raises the question whether these dynamics nevertheless converge to a stable population code in IT, which could offer categorical information to downstream areas, a role typically ascribed to IT. Alternatively, the population dynamics themselves, rather than a final spatial activity pattern, could carry the information relevant to a categorical readout. If the former is the case, decoding from spatial population activity vectors, obtained by averaging over the ‘feedforward’ time window or by selecting the best single time point, should allow for maximal category decoding accuracy. However, if the spatiotemporal trajectories carry categorical information, a decoding model with access to population activity trajectories in IT should perform better (**Fig. 3A**).

We start this comparison by reporting the category decoding performance of the gold-standard linear classification approach that relies on spatial activity patterns. We compare two models, corresponding to the gold standard. The first model represents the single best 10ms window of IT activity, chosen greedily across all time points tested (50 - 240ms). The second model averages the population activity in the classic “feedforward” interval (70-170ms), a simple form of integration. The performances resulted in 23.5% (std=0.8%), and 24.6% (std=1.0%) accuracy, respectively. These “spatial” models were next contrasted with a model that has access to spatiotemporal information. To do so, we trained a small recurrent neural network (RNN) that sequentially receives IT population responses from 20 time points between 50ms and 240ms after stimulus onset (**Fig. 3B**). Using this model, we observed a much improved decoding accuracy of 42.5% (std=0.9%). To rule out the possibility that the additional parameters or nonlinearities of the RNN are the origin of the improvement, we trained the same RNN architecture on the averaged feedforward population activity. At 25.4% (std=0.6%) accuracy, this model performed worse than the RNN trained on population trajectories, and similarly to the linear model trained on the same data. Similarly, shuffling the trajectories in time during training of the RNN largely reduces the model’s accuracy to 30% (std=0.8%). The GRU-based decoding likewise outperformed the other models when using MUAs instead of LFPs (**Fig. S6** & **Fig. S7**). These decoding results underline the possibility that, compared to spatial activity patterns at any single point in time, as implied by a stage-like information transfer, population dynamics provide a richer source of information

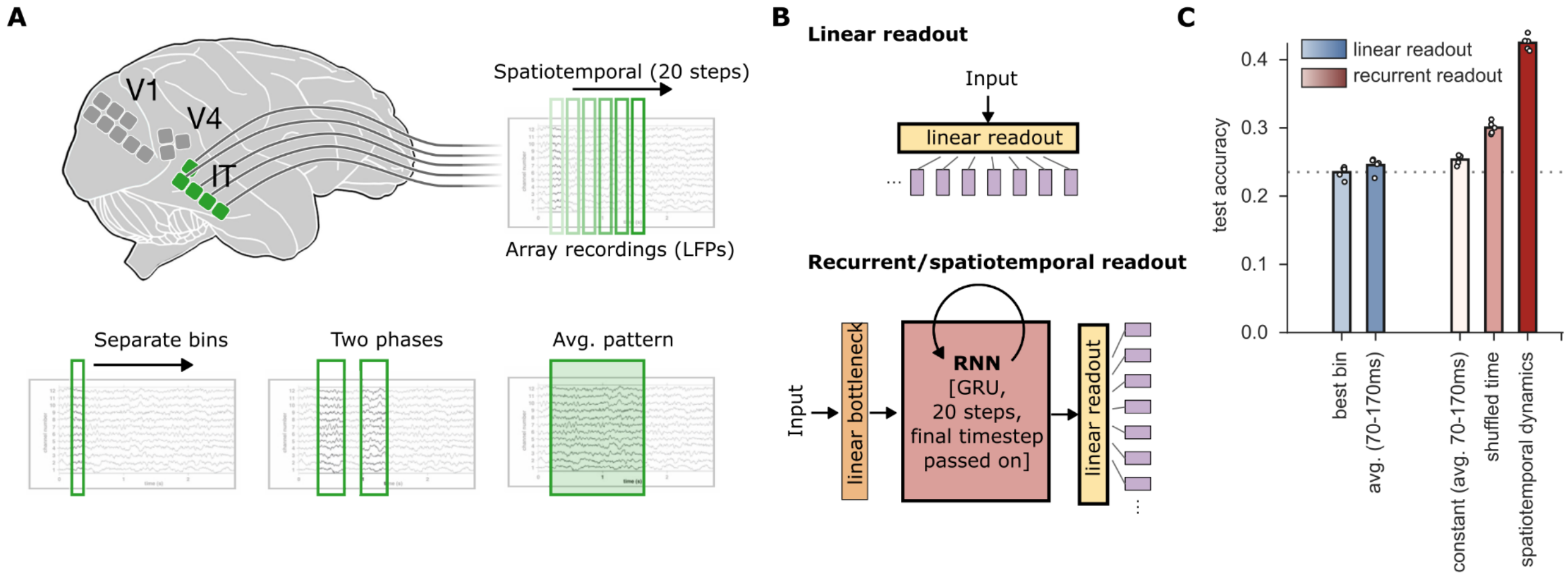

**Fig. 3**: **Trajectories of IT population activity contain categorical information beyond static codes. A,** Models receive data from all IT recording sites in IT from different times after stimulus onset: Linear models are trained on either single 10ms bins, two bins corresponding to the ‘early’ and ‘late’ times identified by inter-area modelling, or data averaged in a large feedforward window spanning 70-170ms. Recurrent models are trained on data from 20 sequential 10ms bins from 50-240ms post stimulus onset. **B,** Decoding models. We compare typical linear decoders (top) to a simple RNN consisting of two linear layers and a GRU cell[11]. **C,** Decoding accuracy improves with access to dynamics. Comparing decoding accuracy between linear models (blue) and recurrent models (red) shows that models with access to trajectories of population activity outperform ‘spatial’ models trained only on spatial patterns. An RNN trained on a constant spatial pattern performs on par with the corresponding linear model ruling out model complexity as an explanation for improved performance. An RNN trained on shuffled trajectories yields decreased accuracy, further emphasising the importance of population trajectories for categorical decoding beyond a set of spatial population activity estimates.

## Discussion

We set out to test the validity of the prevailing view of a single, stage-like feedforward process that transforms spatial response patterns across regions to extract high-level information from retinal input[2]. To this end, we analysed time-resolved, simultaneous recordings from multiple regions of the macaque ventral stream, focusing our efforts on models of inter-area communication. Across all analyses performed, we find evidence for a dynamic, spatiotemporally evolving process that encodes information in its dynamics rather than only spatial response patterns. First, during the time window commonly described as a single "feedforward pass", population activity in V4 was found predictive of downstream IT activity at multiple time points. This suggests that multiple processes unfold within this early window, a view further supported by extensive changes in IT's representational geometry over the same period. Specifically, IT representations transition from similarity with early to late layers of a task-trained ANN, corresponding to a progression from lower- to higher-level feature representations. Second, our RNN decoding approach revealed that IT population dynamics encode rich category information that cannot be decoded from purely spatial estimates of IT population activity. This suggests that rapid visual processing is neither stage-like nor a process that converges onto a stable population code that can then be read out from a single estimate of population activity. Instead, the dynamics of population activity, likely driven by recurrent connectivity[8,12], hold substantial categorical information that can be accessed by downstream recurrent circuits.

This change of view, which covers even the earliest period of visual processing, matches the increasing evidence from modelling and experimental study of different brain networks, suggesting that population dynamics are a relevant computational primitive used by biological and artificial neural networks[13–15]. The dynamics of neural population activity may provide a solution to the need for efficient, flexible computations that adhere to the physical constraints by treating time as a computational resource[16]. This change in perspective has further implications for computational modelling of the ventral stream computations, for which the predominant ANN model class remains feedforward[17]. Beyond providing insights on the processing of visual information in biological systems, this observation can furthermore provide important pointers for the development of future AI systems that may benefit from biological inspiration as an alternative or extension to scale[18–20].

Although the current paradigm collected brain data in response to static stimuli, the emerging view of a dynamic process may benefit studies of more natural experimental paradigms, including active vision[21], or more versatile task settings in which different information content needs to be dynamically extracted over time. Future work also includes causal interventions to study the causal role of proposed dynamic information transfer, as our linear models characterise when and which information in IT is predictable from V4 and thus remain correlational in nature.

Finally, future work also includes the study of which aspects of the neural dynamics are central to the superior decoding performance of our recurrent model for category decoding. Studying which aspects of population activity dynamics are extracted by the model promises further insights into how the brain may use the temporal trajectories for efficient information readout.

## Funding Information

The work was supported by ERC grants 101039524 TIME awarded to TCK (TCK, DA) and 101219949 “SteerMEM" awarded to PP, NWO grant (VI.Veni.222.217) awarded to PP (AM) and DFG RTG grant GRK2340 (PK, DA). Additionally, compute resources for this project are in part funded by the Deutsche Forschungsgemeinschaft (DFG, German Research Foundation), project number 456666331.

## References

1. Felleman, D. J. & Van Essen, D. C. Distributed hierarchical processing in the primate cerebral cortex. *Cereb. Cortex N. Y. N 1991* **1**, 1–47 (1991).

2. DiCarlo, J. J., Zoccolan, D. & Rust, N. C. How Does the Brain Solve Visual Object Recognition? *Neuron* **73**, 415–434 (2012).

3. Hebart, M. N. *et al.* THINGS: A database of 1,854 object concepts and more than 26,000 naturalistic object images. *PLOS ONE* **14**, e0223792 (2019).

4. Thorpe, S., Fize, D. & Marlot, C. Speed of processing in the human visual system. *Nature* **381**, 520–522 (1996).

5. Kar, K., Kubilius, J., Schmidt, K., Issa, E. B. & DiCarlo, J. J. Evidence that recurrent circuits are critical to the ventral stream’s execution of core object recognition behavior. *Nat. Neurosci.* **22**, 974–983 (2019).

6. Hung, C. P., Kreiman, G., Poggio, T. & DiCarlo, J. J. Fast Readout of Object Identity from Macaque Inferior Temporal Cortex. *Science* **310**, 863–866 (2005).

7. Kriegeskorte, N., Mur, M. & Bandettini, P. Representational similarity analysis - connecting the branches of systems neuroscience. *Front. Syst. Neurosci.* **2**, (2008).

8. Kietzmann, T. C. *et al.* Recurrence is required to capture the representational dynamics of the human visual system. *Proc. Natl. Acad. Sci.* **116**, 21854–21863 (2019).

9. He, K., Zhang, X., Ren, S. & Sun, J. Deep Residual Learning for Image Recognition. Preprint at https://doi.org/10.48550/arXiv.1512.03385 (2015).

10. Xiao, W., Vinken, K. & Livingstone, M. Response dynamics in macaque ventral stream recapitulate the visual hierarchy. 2025.11.11.686115 Preprint at https://doi.org/10.1101/2025.11.11.686115 (2025).

11. Cho, K., Merrienboer, B. van, Bahdanau, D. & Bengio, Y. On the Properties of Neural Machine Translation: Encoder-Decoder Approaches. Preprint at https://doi.org/10.48550/arXiv.1409.1259 (2014).

12. Lamme, V. A. F. & Roelfsema, P. R. The distinct modes of vision offered by feedforward and recurrent processing. *Trends Neurosci.* **23**, 571–579 (2000).

13. Oby, E. R. *et al.* Dynamical constraints on neural population activity. *Nat. Neurosci.* **28**, 383–393 (2025).

14. Driscoll, L. N., Shenoy, K. & Sussillo, D. Flexible multitask computation in recurrent networks utilizes shared dynamical motifs. *Nat. Neurosci.* **27**, 1349–1363 (2024).

15. Dunnhofer, M., Wehrheim, M., Ramezanpour, H., Muzellec, S. & Kar, K. Modeling Dynamic Computations in the Primate Ventral Visual Stream. Preprint at https://doi.org/10.48550/arXiv.2601.12258 (2026).

16. Spoerer, C. J., Kietzmann, T. C., Mehrer, J., Charest, I. & Kriegeskorte, N. Recurrent neural networks can explain flexible trading of speed and accuracy in biological vision. *PLOS Comput. Biol.* **16**, e1008215 (2020).

17. Schrimpf, M. *et al.* Brain-Score: Which Artificial Neural Network for Object Recognition is most Brain-Like? 407007 Preprint at https://doi.org/10.1101/407007 (2020).

18. Ali, A., Ahmad, N., De Groot, E., Johannes Van Gerven, M. A. & Kietzmann, T. C. Predictive coding is a consequence of energy efficiency in recurrent neural networks. *Patterns* **3**, 100639 (2022).

19. Lu, Z. *et al.* End-to-end topographic networks as models of cortical map formation and human visual behaviour: moving beyond convolutions. Preprint at https://doi.org/10.48550/arXiv.2308.09431 (2023).

20. Lu, Z., Thorat, S., Cichy, R. M. & Kietzmann, T. C. Adopting a human developmental visual diet yields robust, shape-based AI vision. Preprint at https://doi.org/10.48550/arXiv.2507.03168 (2025).

21. Amme, C. *et al.* Saccade onset, not fixation onset, best explains early responses across the human visual cortex during naturalistic vision. 2024.10.25.620167 Preprint at https://doi.org/10.1101/2024.10.25.620167 (2024).

## Method

### Electrophysiological Recordings

Two macaque monkeys (males, 7 and 6 years old) participated in the experiments. They were socially housed in stable pairs in a specialized primate facility with natural daylight, controlled humidity and temperature. The homecage was a large floor-to-ceiling cage which allowed natural climbing and swinging behavior. The cage had a solid floor, covered with sawdust and was enriched with toys and foraging items. Their diet consisted of monkey chow, supplemented with fresh fruit. Their access to fluid was controlled, according to a carefully designed regime for fluid uptake. During weekdays, the animals received water or diluted fruit juice in the experimental set-up upon completed trials. We ensured that the animals drank sufficient fluid in the set-up and supplemented the animals with extra fluid after the recording session if they did not drink enough. During the weekend, they received a full bottle of water (700–940 ml per day) in the home cage. The animals were regularly checked by veterinary staff and animal caretakers and their weight and general appearance were recorded daily in an electronic logbook during fluid-control periods. All procedures complied with the NIH Guide for Care and Use of Laboratory Animals and were approved by the ‘‘Centrale Commissie Dierproeven’’ of the Netherlands.

The two macaque monkeys were implanted with 16 and 15 64-channel Utah arrays, respectively. Arrays were implanted across the ventral stream in V1 (8 & 7 arrays), V4 (3 & 4 arrays), and IT (5 and 4 arrays). Experimental setups are described in Papale et al.[1], where they do not deviate from the method used in the present study. Raw signal was recorded simultaneously from all electrode arrays at 30kHz and filtered between 0.1 - 200hz with a second-order Butterworth filter applied using a zero-phase filtering routine (effectively implementing a fourth-order Butterworth filter). Additionally, notch filters were applied at 50Hz and the first four harmonics with a notch width of +- 5Hz, using the same filter design and routine. After filtering, data were downsampled to 1kHz. For both monkeys, the experiment was repeated in six sessions each over three months. For monkeyN, two sessions were excluded from the analyses due to issues with the recordings. During each completed experimental session, all 1000 stimuli were shown three times. In some sessions, not all trials were completed, leading to a total of 6 sessions with 16.478 trials for monkeyF and 4 sessions with 11.998 trials for monkeyN that are included in the analyses.

### Channel Rejection

To exclude noisy electrodes and recording sites that do not respond to visual stimulation, we compute a d-prime measure for each electrode, quantifying response amplitude relative to baseline. D-prime for an electrode is computed separately for each session and category. Within each session, we record the largest d-prime value over all 100 categories since, especially in IT, many recording sites respond strongly only to a subset of categories. Computing d-prime jointly for trials from all categories would systematically lower d-prime estimates for recording sites that selectively respond only to few stimulus categories. An electrode is included if this d-prime value exceeds a threshold of 1.5 for all sessions. This cutoff was chosen by visual inspection of d-prime distributions computed from recordings known to have no signal, where almost no channels exceed this threshold.

For one electrode, session, and category, d-prime is computed as follows: We define a baseline interval as -100-0ms relative to stimulus onset, as well as a ‘signal’ interval containing the same number of samples. This window is staggered over ROIs to account for response onset latencies and is defined as 25-125ms in V1, 50-150ms in V4, and 75-175ms in IT. For all included trials, we subtract the mean signal computed jointly over the baseline period of all trials. Since electrodes show positive and negative signal deflections in response to visual stimuli, we then compute for each trial the mean of the absolute signal during both the baseline periods denoted $b$ and during signal periods denoted s. Finally, we compute the mean and standard deviation over these measures across trials and denote them $\mu_b$ and ${\sigma^2}_b$ for the baseline period and $\mu_s$ and ${\sigma^2}_s$ for the signal period. D-prime is then computed as:

$$d' = \frac{\mu_s - \mu_b}{\frac{1}{2}\sqrt{{\sigma^2}_b^{\,2} + {\sigma^2}_s^{\,2}}}$$

### Additional Preprocessing

Since all analyses are concerned with trial-by-trial variation in neural responses, we standardize the responses for each channel and time point by computing the z-score over trials separately for each session. This additionally accounts for potential session-wise differences in signal amplitude. To increase the signal-to-noise ratio for our analyses, we subsequently average recordings in sliding 10ms bins and subsequently re-standardize data jointly for data pooled from all included sessions for each monkey. Our choice of relatively short bins is motivated by the consideration that 10ms bins centered on reported time points for our inter-area analyses ensure that bins for predictor and predicted time points never overlap for delays exceeding 10ms, while we expect to find conduction delays between regions on the order of 20ms or longer.

### Experimental Procedure

Stimuli were presented in a blocked rapid serial visual presentation (RSVP) paradigm. During each block, monkeys had to fixate on a fixation point to obtain a juice reward. Each block started with a fixation point displayed on a gray screen for 300ms, after which 4 stimuli were shown for 250ms each, with jittered inter-stimulus intervals between 150-250ms. If a monkey broke fixation, the block ended, and the remaining stimuli were shown during the next block. Stimuli from this block that were completed before fixation was broken were not repeated. In each completed session, each of 1000 stimuli was presented 3 times.

### Stimuli

The stimuli used in the experiment are a manually curated subset of the THINGS[2] dataset. Out of the available 1854 object concepts, we manually chose 100, with 10 images each. Concepts and images were selected to be a good sample of the stimulus diversity of the full dataset, while also creating a balanced hierarchy of stimulus categories and limiting the total number of stimuli to an amount that can be repeated multiple times in each session. An additional criterion for inclusion of any object concept was that at least 10 images were available that prominently feature the object concept, and minimize the presence of other objects in the frame.

### Oracle Correlation

Oracle correlation is computed separately for each electrode, jointly for data from all sessions. For each stimulus, we select one trial at random, obtaining a 1000-dimensional vector. A second vector is computed by averaging the responses from the remaining trials for each stimulus. We estimate one sample of the oracle correlation as the Pearson correlation between these two vectors. For each electrode, we repeat this procedure for 30 random samples and report the mean. Thin lines in Fig. 1C and E show oracle correlation over time for each electrode, with bold lines showing the average over all channels within an ROI.

### Category Decoding

For category decoding (Fig. 1D), we fit Ridge Classifiers implemented with sklearn[3]. Since our dataset contains the z-scored trial-by-trial deviation from the average response, the data is already centered, and we fit all models without intercept. To account for any class imbalance as a result of recording sessions that were not completed, the loss is 'balanced' - applying a weighting to account for the number of training samples per class. We fit one model every 5ms, choosing the regularization parameter alpha by cross-validation for every time bin. Alphas are chosen from 128 log-spaced values ranging from $10^0$ - $10^8$. Cross-validation is performed over 50 folds, where for each fold we randomly sample one stimulus for each category and assign all corresponding trials as the validation split. For each split, this assigns approximately 10% of trials to the validation set and ensures that category-level decoding is tested on responses to stimuli that are not used to fit the classifier[4]. Reported accuracies are averages over cross-validation splits.

### Inter-area modeling

We model inter-area information transfer with a set of linear models in the 'Granger-causality'[5] framework. To estimate the extent to which activity in V4 is predictive of activity in a target array in IT for a specified time $t$ and delay $d$, we fit two linear models.

The first predicts activity in the target array $Y$ at time $t$ from activity measured at the same electrodes at the earlier time $t - d$.

$$\hat{Y}_t = W \cdot Y_{t-d}$$

The second model has activity measured at all electrodes in V4 as an additional set of predictors $X$.

$$\hat{Y}_t = W_r \cdot Y_{t-d} + W_i \cdot X_{t-d}$$

The contribution of V4 for predicting the activity at a target array in IT is the increase in explained variance (measured with the coefficient of determination) in the second model over the first. The rationale behind this approach is that for our analyses, it is important to know 'when' information in V4 is newly predictive of activity in IT - as an estimate of the potential timing of information transfer. Since neural measurements are autocorrelated in time, we include IT's past as a predictor to capture variance that is already present in the target electrodes at earlier time points. We expect that this yields a more conservative estimate of the contribution of information from V4. Additionally, this modeling choice can also account for trial-to-trial variation that affects measurements at all sensors simultaneously. Importantly, we do not explicitly orthogonalize the two sets of predictors.

Both models are implemented as ridge regressors without intercept. Regularization parameter alpha is selected separately for each target electrode and chosen from 2048 log-spaced values in the range $10^{-3}$ to $10^8$.

Since we perform a number of analyses on predictions made using these models, a primary concern for cross-validation is to leave sufficient data that can be analysed without reusing trials used to fit the models. This is especially important for representational similarity analysis, for which we need sufficient repeated trials for all stimuli to compute dissimilarity matrices. To achieve this, trials are split into three disjoint folds, stratifying over sessions and stimuli. We use each split as the training data for one model, and choose regularization parameters alpha using efficient leave-one-out cross-validation on each train split. Reported model performance is computed jointly over the remaining two-thirds of the data that form the test split for each fold. Separate inter-area models are fit to predict each target array in IT, from all channels in V4, sharing the same train-test splits.

We fit models for a large grid of target times (0-300ms post stimulus onset) and delays (2-80ms), with a stride of 2ms for both time and delay. This yields a total of 6.040 sets of models.

To investigate the subspaces read out from V4 to predict a target array in IT at different times, we focus our analyses on the weight matrix $W_i$ that maps activity from all channels in V4 to activity in a target array in IT. Specifically, we select models for two (time, delay) pairs for further analysis and refer to them as the 'early' and 'late' models throughout:

$$\hat{Y}_t = W_i \cdot X_{t-d}$$

### Comparing model similarity

To assess the similarity of models fit for different times and delays (Fig. 1G), we extract weight matrices $W_i$ from models for all (time, delay) pairs, target arrays in IT, and cross-validation folds. The distance between the selected ('early' or 'late') model and all other models is computed using the cosine distance between each pair of flattened weight matrices. Results presented in Fig. 1G are then averaged over folds and models for different target arrays in IT.

### Predictions with reduced rank regression

Previous work modeling inter-area interactions between regions of the ventral stream[6] shows that interactions between regions are low-rank compared to lateral interactions within a neural population. To explicitly model that only a subspace of V4 population activity is 'read out' to predict activity in each array in IT, we perform rank-reduction on each $W_i$ using the reduced-rank regression framework[7], similar to previous work[6]. In addition to analyzing predicted activity in IT, this also allows for complementary analyses of the subspaces read out from V4.

To obtain the best reduced-rank approximation of the weight matrix $W_i$, we obtain model predictions $\hat{Y}_t$ for the train set of the model and compute their singular value decomposition. This gives a set of singular values and corresponding singular vectors. These are used to iteratively reduce the rank of the matrix $W_i$, starting with directions with the smallest singular vectors. The final rank of the reduced-rank version of $W_i$ is chosen by cross-validation as the rank that yields the best average validation accuracy over the three folds. Whenever an analysis in the following is concerned with predicted time courses, these predictions have been made using the reduced rank version of $W_i$.

### Comparing the representational geometry of predictions by different models

To compare the representational geometry of predictions of IT obtained from either an 'early' or 'late' model of information transfer from V4 to IT, we obtain predictions from both an 'early' and a 'late' model for the time and delay at which the models were fit - using only trials from the validation split for each model. Subsequently, representational dissimilarity matrices (RDMs) are computed for all 'predicted' data sets. RDMs are then averaged over the three folds. To compare geometries across predictions for all arrays in IT and both time points, we compute the rank order correlation (Spearman's $\rho$) between all RDMs.

To test whether RDMs predicted by 'early' and 'late' models for all recording sites in IT (5 for monkeyF, 4 for monkeyN) are more similar within the same time, as compared to across time, we perform a permutation test against a model RDM. The model RDM is constructed to have 0 distances between RDMs for all arrays predicted at the same time, and 1 for all distances between RDMs predicted using models from different time points. The 'true' RDM is obtained by converting the matrix of rank-order correlations for all pairs of RDMs $S$ to a distance matrix $D$: $D = 1 - S$. The test statistic is again the rank order correlation (Spearman's $\rho$) between the second-order RDM over predictions $D$ and the model RDM. We then compute the same test statistic for 10.000 joint permutations of the rows and columns of $D$ and report the fraction of test statistics in our sample that are larger than for the unshuffled RDM.

### Unique explained variance of predicted trajectories

To compare the extent to which predictions made using 'early' and 'late' models align with the representational geometry of population activity measured from arrays in IT over the whole trial time course, we extend our representational similarity analysis (RSA) to predictions of whole IT trajectories from trajectories in V4. We then compute two sets of RDM movies from the predictions made using either 'early' or 'late' models. RDMs are computed separately for each fold and then averaged over folds.

For each time point and array in IT, we fit GLMs using nonnegative least squares to predict the RDM computed from recordings in this array and time from two predictors: the predicted RDMs for the same time point made using the corresponding 'early' and 'late' models. We compute the unique explained variance of a predictor by refitting the GLM, excluding this predictor and taking the difference in explained variance between the model with all predictors and the model excluding the predictor of interest.

Finally, to find time points where either the 'early' or 'late' model explains significantly more unique variance than the other, we perform a cluster permutation test comparing unique explained variance trajectories, normalized by the explained variance of the model containing both predictors, for the two models against each other. We perform 1024 permutations and use a one-sided F-statistic as the test statistic.

## Characterizing predictions of IT activity with representational similarity analysis

### Neural RDMs

For representational similarity analysis[8], representational distance matrices (RDMs) are computed at the stimulus level. For neural data, all RDMs are computed using correlation distance. To account for differing numbers of trials for each stimulus condition due to missing trials, we compute RDMs from a fixed-size sample of trials for each stimulus and average over sample RDMs. This ensures that distances are not biased by the number of trials available for each condition. RDMs are averaged over 1000 samples. For each sample, we randomly select without replacement 10 trials per stimulus for monkeyF, and 7 trials for monkeyN to account for the different dataset sizes for the two monkeys.

### DNN RDMs

We use TIMM[9] to extract embeddings for all 1000 stimuli from each convolutional and linear layer (before nonlinearity) of a Resnet18[10] trained on ImageNet. For convolutional layers, embeddings are flattened over the height, width, and channel axes to obtain 1D embeddings. To account for the large number of features, we then reduce the embeddings from each layer to 1000 dimensions. Differently to RDMs for neural data, dissimilarity matrices for ResNet embeddings are computed using cosine distance since this metric is invariant to rotations of the coordinate system that may result from applying PCA[11].

### Correlating RDMs from predicted time courses with model RDMs

To characterize the representational content of predicted time courses for each array in IT, we compute the Pearson correlation between each model RDM obtained from ResNet18[10] and predicted RDM time course in IT. Correlations are computed separately for RDM timecourses for each target array, model, and cross-validation fold and then averaged for visualization in Fig. 2b.

## Recurrent Decoders

When comparing linear models to recurrent neural network (RNN) based decoders, a similar setup to the above decoding analysis is used. One notable exception is that for these analyses reported in Fig. 3C, cross-validation is performed over 5 folds only, to account for the higher computational cost of training RNN decoders. Additionally, before fitting each linear decoder, the data are transformed to a set of orthogonal predictors by fitting PCA to the training set, without reducing the number of components.

To compare different decoders, we create a number of datasets, selecting data from different time points from all included recording sites in IT. First, to estimate the decoding accuracy for the best single time point, we fit linear models for 20 time bins, starting 50ms up to 240ms after stimulus onset, in 10ms steps. We report accuracy for the best time point, averaging over cross-validation folds. Second, for decoding from the "feedforward average" signal, we fit models after averaging responses in time between 70-170ms post stimulus onset. Finally, to jointly decode from the two time points identified to be predictable from V4 and having qualitatively different representational geometry, we concatenate measurements for two time points corresponding to the 'early' and 'late' model of the inter-area analyses - resulting in a model with two predictors per channel in IT.

The recurrent decoder is an RNN implemented in PyTorch[12]. The architecture consists of a linear 'bottleneck' layer projecting data from all channels in IT to 32 dimensions, followed by layer normalisation[13] and dropout[14] (p=0.3). Embeddings from each time point are then fed through a layer of 64 'gated recurrent units' (GRU)[15]. Layer normalization and dropout (p=0.3) are applied to the output of this recurrent block after the last time step. A final linear layer and softmax nonlinearity map this normalized output of the recurrent block after the last time step to a probability distribution over the 100 categories.

All recurrent networks are optimized to minimize crossentropy loss with AdamW[16] (learning rate = 0.001, $\beta_1$= 0.9, $\beta_2$ = 0.999, $\epsilon$ = $10^{-8}$, weight decay $l2$ = 0.01). For each cross-validation fold, we perform early stopping based on validation performance. RNNs are unrolled for 20 time steps, sequentially receiving data from 50-240ms post stimulus onset in 10ms steps as described above ('spatiotemporal dynamics', Fig. 3C).

We additionally train the same RNNs on a number of control conditions. One set of RNNs is trained on the 'feedforward average' dataset used to train the corresponding linear decoders, unrolling the network for 20 steps during which the input drive to the network is constant ('constant', Fig. 3C). An additional set of RNNs is trained on trajectories shuffled along the time axis. Randomization is applied separately each time a sample is used for training. Validation performance is computed on unshuffled trajectories. Finally, one set of RNNs is trained only on the signal change from one time point to the next ('pattern change', Fig. 3C). The dataset for this model is created by computing the difference between data from adjacent time bins, resulting in a trajectory containing 19 time points. Notably, as opposed to all other networks, which are trained on data z-scored for each time bin, signal change here is computed on the time courses prior to z-scoring in order to preserve information on the magnitude of signal change at different times relative to stimulus onset. After temporal differencing, the data are standardized by z-scoring signal change for each channel jointly over the trial and time axis. This z-score is computed separately for data from each recording session, and mean and standard deviation are estimated on the train split and applied to trials from the validation set.

## References

1. Papale, P., Wang, F., Self, M. W. & Roelfsema, P. R. An extensive dataset of spiking activity to reveal the syntax of the ventral stream. *Neuron* **0**, (2025).

2. Hebart, M. N. *et al.* THINGS: A database of 1,854 object concepts and more than 26,000 naturalistic object images. *PLOS ONE* **14**, e0223792 (2019).

3. Pedregosa, F. *et al.* Scikit-learn: Machine Learning in Python. *J. Mach. Learn. Res.* **12**, 2825–2830 (2011).

4. Kilgallen, J., Pearlmutter, B. & Siskind, J. *The Repeated-Stimulus Confound in Electroencephalography*. (2025). doi:10.48550/arXiv.2508.00531.

5. Granger, C. W. J. Investigating Causal Relations by Econometric Models and Cross-spectral Methods. https://www.jstor.org/stable/1912791?seq=1 (1969).

6. Semedo, J. D., Zandvakili, A., Machens, C. K., Yu, B. M. & Kohn, A. Cortical Areas Interact through a Communication Subspace. *Neuron* **102**, 249-259.e4 (2019).

7. Izenman, A. J. Reduced-rank regression for the multivariate linear model. *J. Multivar. Anal.* **5**, 248–264 (1975).

8. Kriegeskorte, N., Mur, M. & Bandettini, P. Representational similarity analysis - connecting the branches of systems neuroscience. *Front. Syst. Neurosci.* **2**, (2008).

9. Wightman, R. PyTorch Image Models. *GitHub repository* (2019) doi:10.5281/zenodo.4414861.

10. He, K., Zhang, X., Ren, S. & Sun, J. Deep residual learning for image recognition. in *Proceedings of the IEEE conference on computer vision and pattern recognition* 770–778 (2016).

11. Mehrer, J., Spoerer, C. J., Kriegeskorte, N. & Kietzmann, T. C. Individual differences among deep neural network models. *Nat. Commun.* **11**, 5725 (2020).

12. Paszke, A. *et al.* PyTorch: An Imperative Style, High-Performance Deep Learning Library. in *Advances in Neural Information Processing Systems* vol. 32 (Curran Associates, Inc., 2019).

13. Ba, J. L., Kiros, J. R. & Hinton, G. E. Layer Normalization. Preprint at https://doi.org/10.48550/arXiv.1607.06450 (2016).

14. Srivastava, N., Hinton, G., Krizhevsky, A., Sutskever, I. & Salakhutdinov, R. Dropout: A Simple Way to Prevent Neural Networks from Overfitting. *J. Mach. Learn. Res.* **15**, 1929–1958 (2014).

15. Cho, K., Merrienboer, B. van, Bahdanau, D. & Bengio, Y. On the Properties of Neural Machine Translation: Encoder-Decoder Approaches. Preprint at https://doi.org/10.48550/arXiv.1409.1259 (2014).

16. Loshchilov, I. & Hutter, F. Decoupled Weight Decay Regularization. Preprint at https://doi.org/10.48550/arXiv.1711.05101 (2019).

## Supplementary Materials

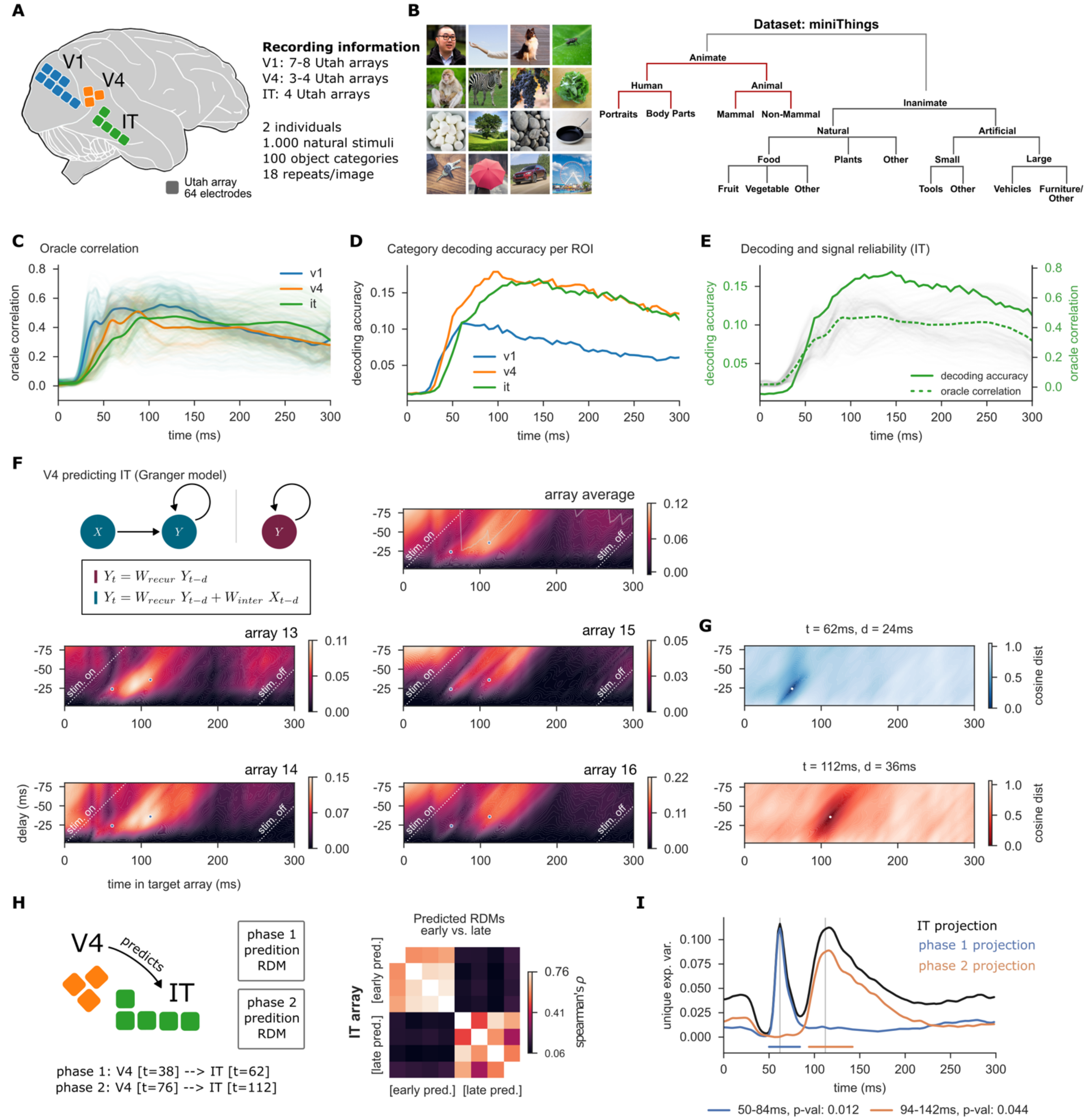

**Fig. S1**: **Predicting information transfer from V4 to IT during feedforward processing of visual information. Data for monkeyN. A,** Two macaque monkeys were implanted with multiple electrode arrays across V1, V4, and IT. LFPs were recorded while monkeys passively viewed natural stimuli. **B,** Stimuli are a curated subset of THINGS[1] from 100 categories with 10 exemplars each. Categories form a hierarchy of broad to specific 'super categories', sampling the diversity of THINGS. **C - E,** ROI-wise analyses resolved in time. The onset of stimulus related information measured with oracle correlation and categorical information measured with linear probes are both staggered in time with increasing latency from V1 to IT. **F,** Modeling inter-area information transfer. Neural activity at each of 5 arrays implanted in IT can be predicted from arrays in V4 at multiple times after stimulus onset. For each target array, there is an 'early' and 'late' cluster of timepoints where V4 predicts IT. One set of models per cluster is selected for further analysis: The 'early' models predicting IT at 74ms from V4 at 46ms are indicated with green points. The 'late' models predicting arrays in IT at 120ms from V4 at 80ms are indicated with blue points. **G,** Similarity of models over time. Surfaces show cosine distance between the weight matrices of the 'early' (top) and 'late' (bottom) models compared to models from all other times. There are separate clusters of similar models 'early' and 'late' that are different from models in the other cluster. Distances are computed separately between models for each target array and surfaces shown here are averages over arrays. **H,** representational similarity of 'early' and 'late' predictions of IT. Representational geometries of predicted signals in arrays in IT are similar for all models from the same time, but differ between 'early' and 'late' times. **I,** 'Early' and 'late' models uniquely predict representational geometry in IT in intervals around times at which they were fit.

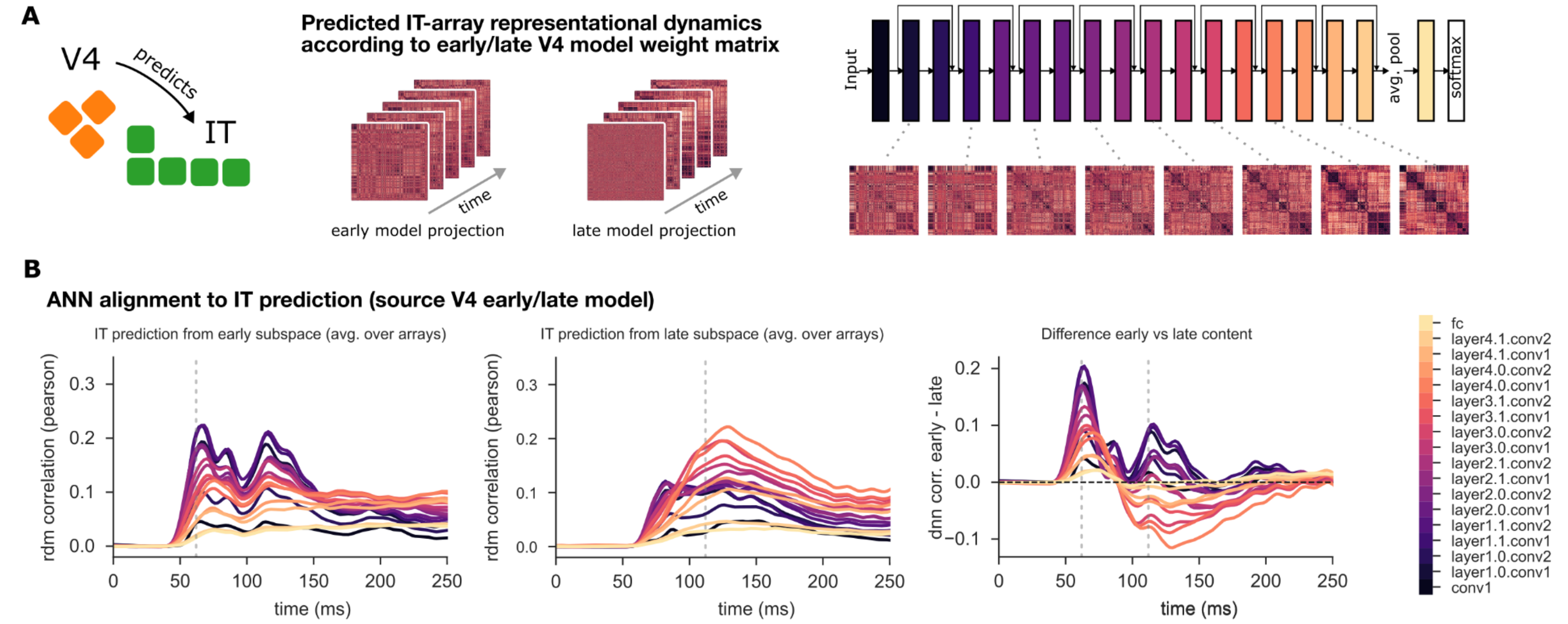

**Fig. S2**: **Characterizing representational geometry of early and late predictions of recording sites in IT with RSA. Data for monkeyN. A,** Two sets of RDM movies for arrays in IT are predicted from V4 time courses using the 'early' and 'late' inter-area models selected for analysis (left). RDMs are extracted from multiple layers of ResNet18[2] as models for 'low-level' geometries (early layers) and 'categorical' geometries (late layers). **B,** Characterization of predicted representational geometries over time. RDM trajectories predicted by 'early' (left) and 'late' (middle) models of inter-area interaction are correlated with model RDMs extracted from ResNet18. Predicted time courses from 'early' models correlate most highly with early ResNet layers, while time courses predicted from 'late' models show increased correlation with later layers. Dashed vertical lines indicate times for which models were fit. Taking the difference between the two panels (right) indicates that representational geometries predicted from 'early' and 'late' models are qualitatively different for late times as well, suggesting that differences in population activity falling onto either readout from V4 persist for the duration of stimulus presentation.

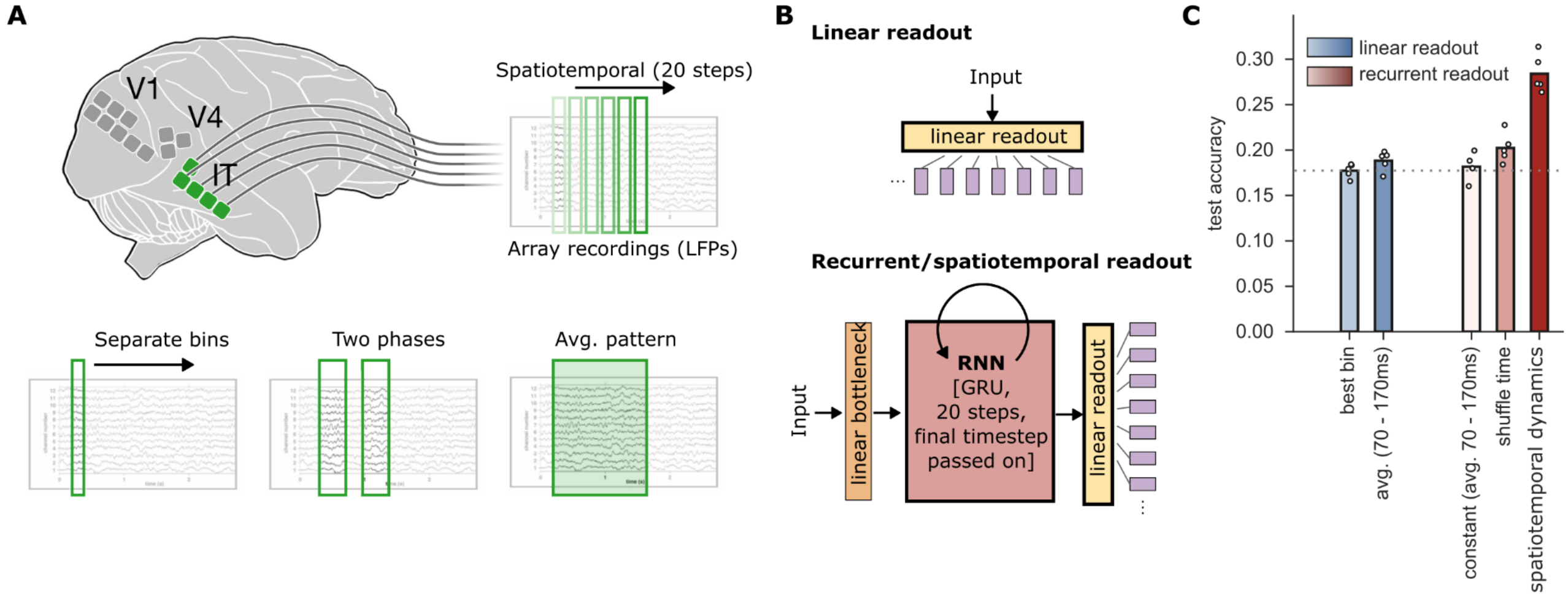

**Fig. S3**: **Trajectories of IT population activity contain categorical information beyond static codes. Data for monkeyN. A,** Models receive data from all IT recording sites in IT from different times after stimulus onset: Linear models are trained on either single 10ms bins, two bins corresponding to the 'early' and 'late' times identified by inter-area modeling, or data averaged in a large feedforward window spanning 70-170ms. Recurrent models are trained on data from 20 sequential 10ms bins from 50-240ms post stimulus onset. **B,** Decoding models. We compare typical linear decoders (top) to a simple RNN consisting of two linear layers and a GRU cell[3]. **C,** Decoding accuracy improves with access to dynamics. Comparing decoding accuracy between linear models (blue) and recurrent models (red) shows that models with access to trajectories of population activity outperform 'spatial' models trained only on spatial patterns. An RNN trained on a constant spatial pattern performs on par with the corresponding linear model, ruling out model complexity as an explanation for improved performance. An RNN trained on shuffled trajectories yields decreased accuracy, further emphasising the importance of population trajectories for categorical decoding beyond a set of spatial population activity estimates.

## Analyses of multi-unit activity (MUA)

While the main results presented here are derived from analyses of the LFP, selected analyses were repeated using the MUA. MUA were obtained using the same preprocessing steps as described by Papale et al[4]: Raw data recorded at 30kHz were band-pass filtered between 500Hz - 5kHz with a second-order Butterworth filter using a zero-phase filtering routine. Filtered data were rectified and smoothed with a low-pass filter at 200Hz. Additional filters were applied at 50 and 60Hz and the respective harmonics. Finally, data were downsampled to a sampling rate of 1kHz.

## Included recording sites for LFP and MUA

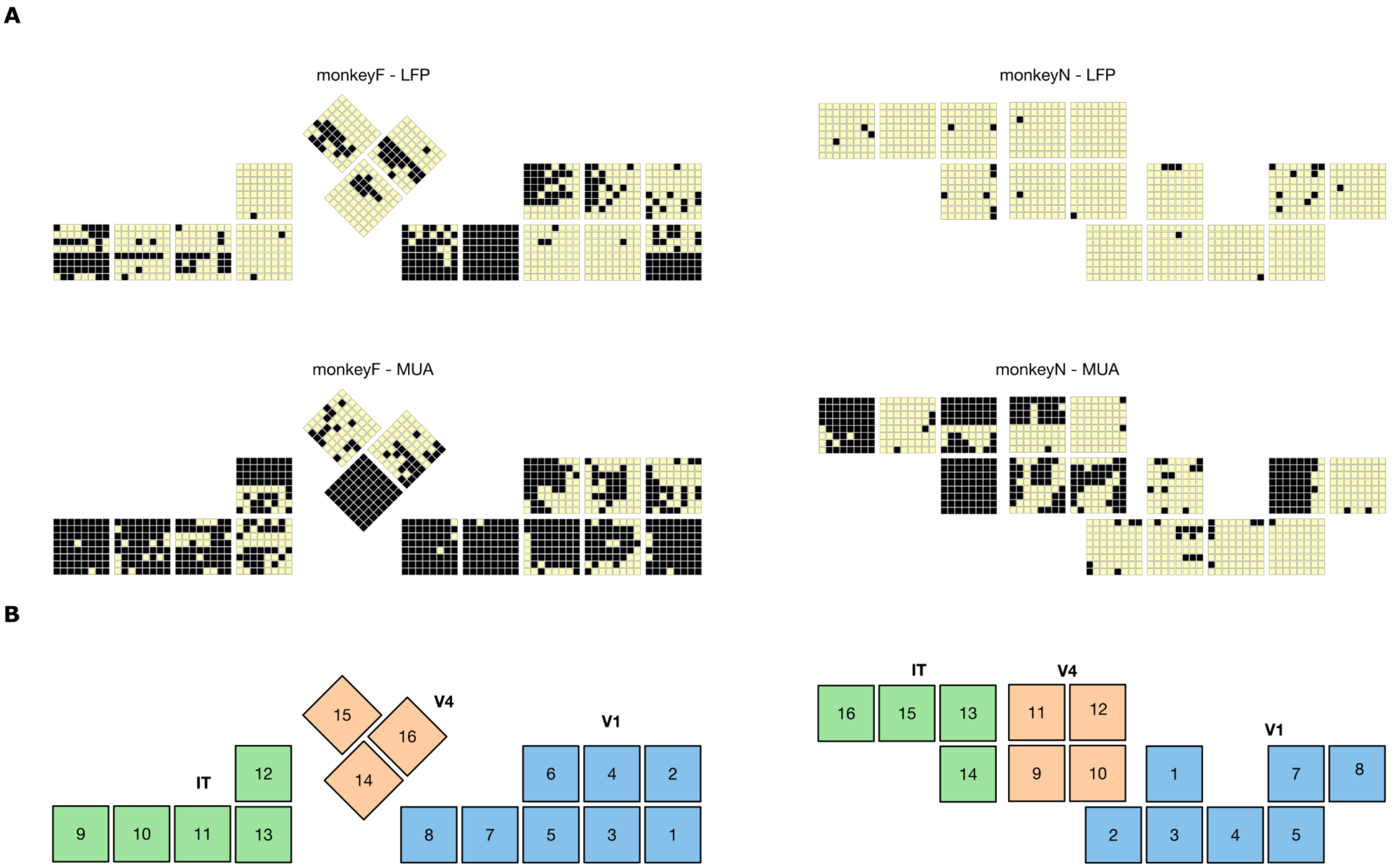

**Fig. S4: included channels for LFP and MUA. A,** Included channels for analyses of the LFP (top), and multi-unit activity (MUA) (bottom). Included channels are brightly colored, and excluded channels are shown in black. Channels are included if they have d-prime > 1.5 for all included sessions. **B,** Schema of array locations for both monkeys. Each rectangle is one Utah array; numbers indicate the array id used to reference recording sites in figures.

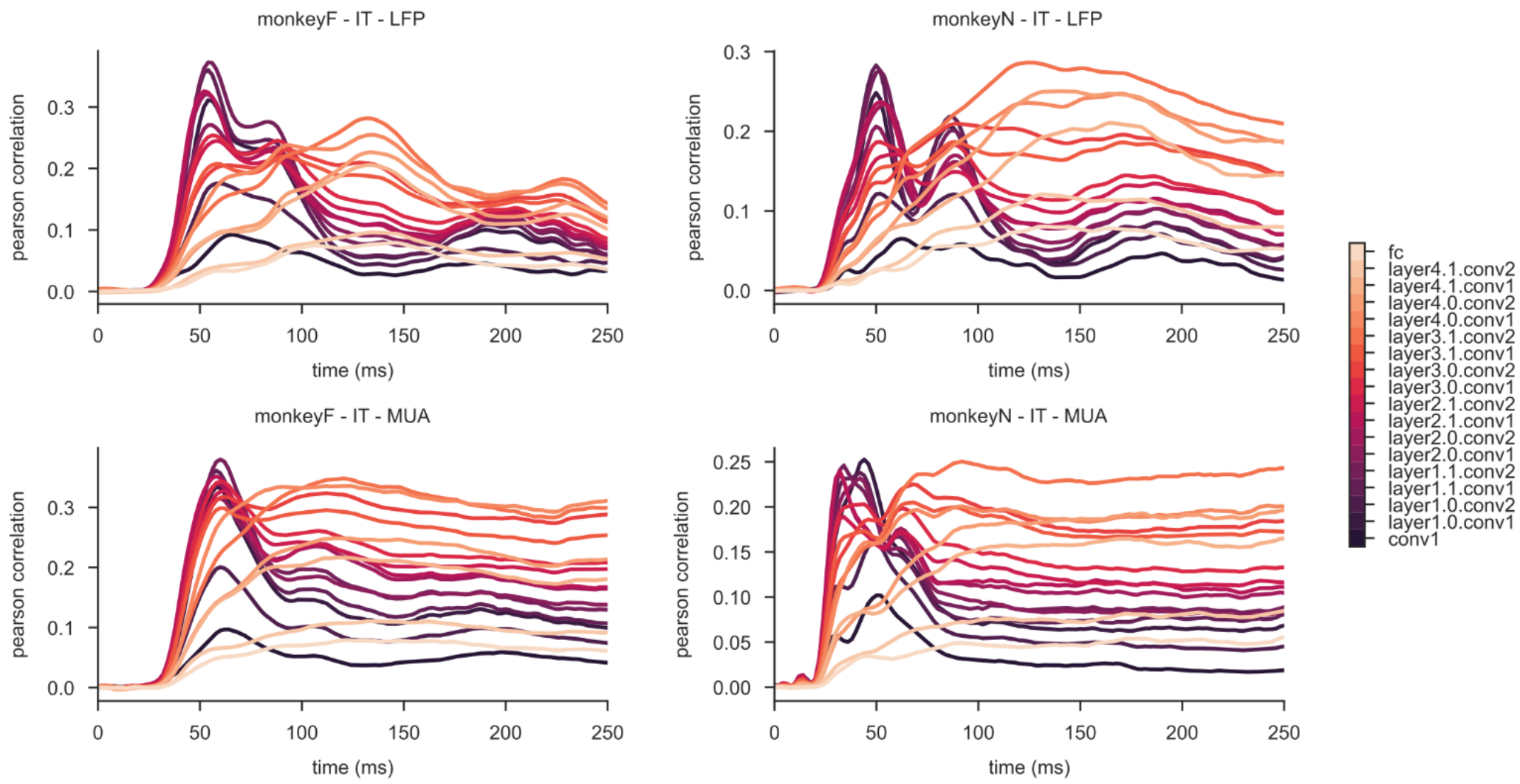

**Fig S5: RDM correlations between IT timecourses and ResNet18 layers.** In addition to the RDM correlations based on predicted signals for arrays in IT, we have computed correlations between model RDMs extracted from ResNet18 and the RDM movies from the recorded signals in IT. We repeated this analysis for both the LFP, as used for the main analyses, and the MUA. Correlations are shown both for neural RDM movies computed from the local field potentials (top) and the multi-unit activity (bottom). For both monkeyF (left) and monkeyN (right), and both types of neural data, there is a transient period early during which representational geometry in IT is most similar to early ResNet18[2] layers, indicating a representational geometry driven primarily by low-level features. At later times, representational geometry in IT correlates best with later ResNet18 layers, indicating a geometry driven by high-level features.

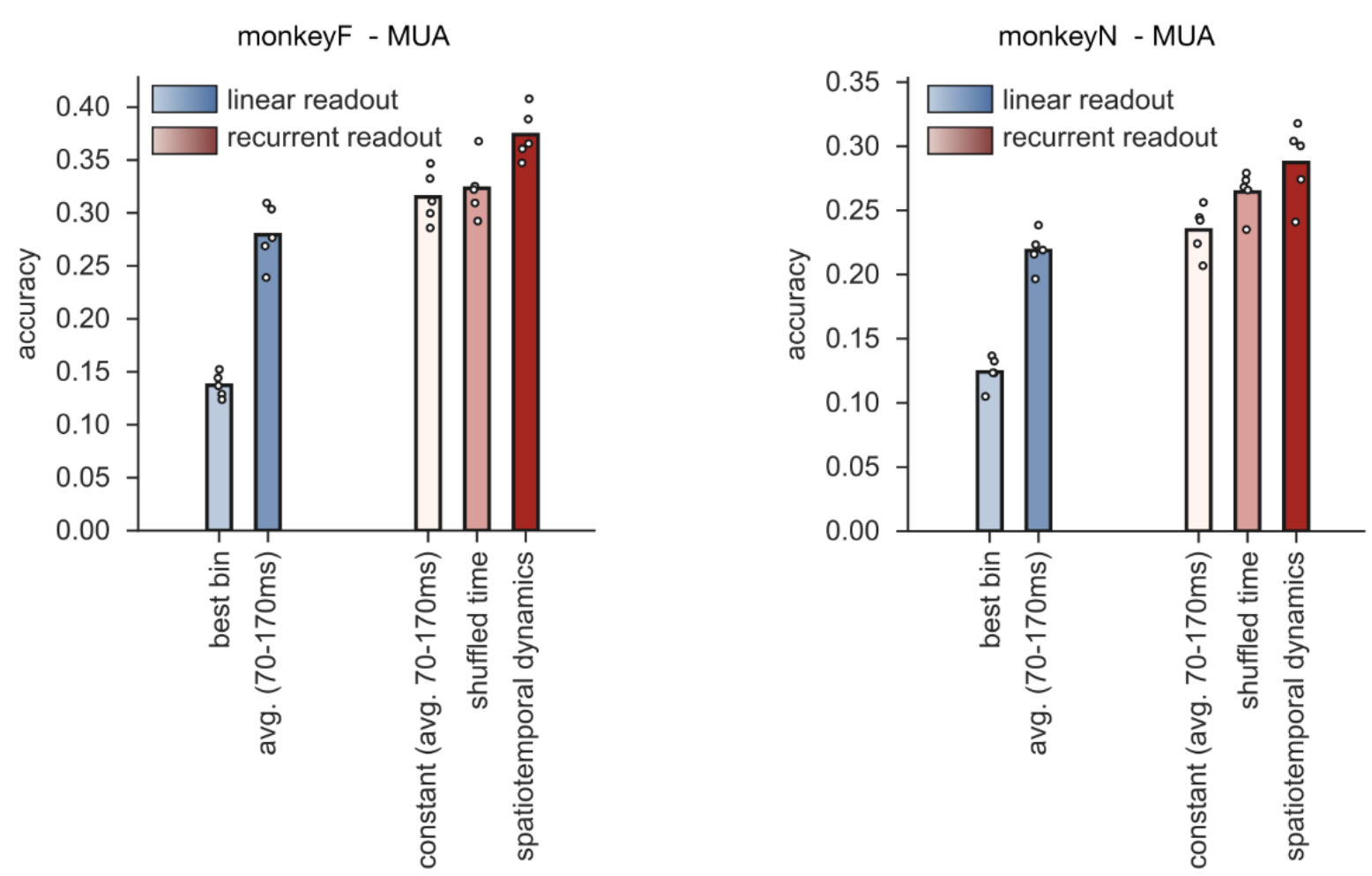

**Fig. S6: Comparison of decoders for multi-unit activity.** Decoding analyses presented in Fig. 3 were repeated for both monkeys using MUA instead of LFP. As before, decoding accuracy improves with access to dynamics. Notably, in contrast to the results of the analyses using LFP (Fig. 3), both the linear model fit on the 'feedforward average', as well as the RNN trained on the same static input, outperform a linear model trained on the best 10ms bin. However, both models trained on spatial patterns are still outperformed by the RNN with access to the full trajectories.

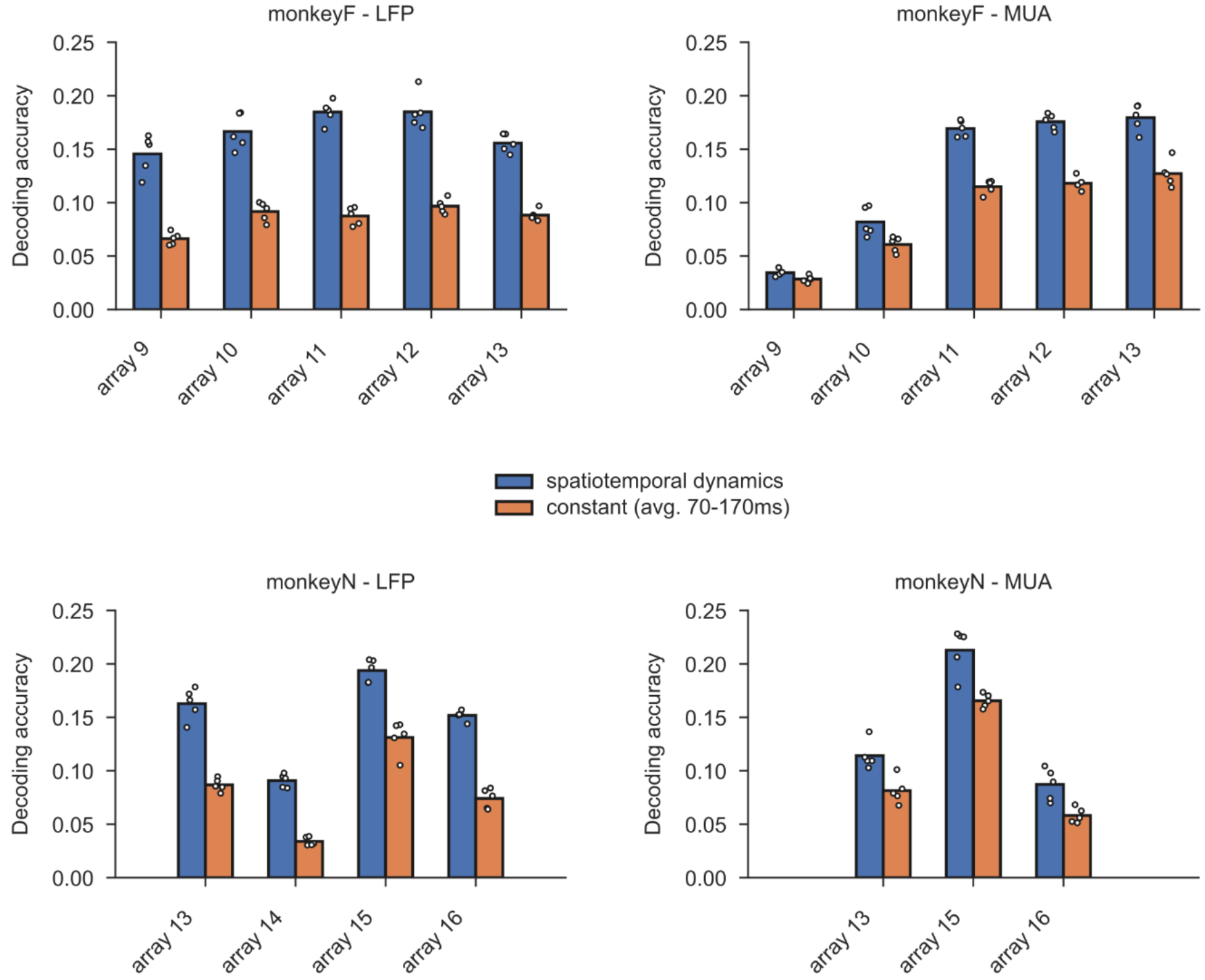

**Fig. S7: array-wise recurrent decoding.** We hypothesized that the increased performance for decoding from the feedforward averaged signals (70-170ms) for MUA over LFPs may be related to the increased spatial resolution of MUA, which might decrease the dependence of the decoder on temporal information. This may be especially relevant for this dataset, because electrodes from multiple Utah arrays implanted across IT cover a large anatomical extent. Results from the same monkeys show staggered response onsets as a function of anatomical location in IT[4]. As a consequence of this, while there may not exist a single 10ms window that covers the most informative signal at all electrodes in IT simultaneously, the average response over a 100ms window may already contain some information encoded in the full dynamics.

To test this hypothesis, we repeat the decoding analysis (Fig. 3) for anatomically constrained sets of electrodes. RNNs for category decoding are trained separately for each array in IT, and both the LFP and MUA. The architecture of the RNN is similar to the model used for the analysis described in the main text, with the exception that the size of the model is reduced to account for the reduced number of recording sites included for each model (8-dimensional bottleneck, 16-dimensional GRU block). We compare the performance of the RNN trained with access to the full neural trajectories to the same network trained on the static feedforward averages (70-170ms) as before.

For all arrays and both LFP and MUA, we observe that the models trained on the population dynamics outperform models trained on static feedforward averages of population activity. From this, we conclude that, especially from the perspective of a neural population that receives input from a spatially constrained neural population in IT, population dynamics carry information relevant to category decoding that is not accessible from a single spatial estimate of population activity.

## References

1. Hebart, M. N. *et al.* THINGS: A database of 1,854 object concepts and more than 26,000 naturalistic object images. *PLOS ONE* **14**, e0223792 (2019).

2. He, K., Zhang, X., Ren, S. & Sun, J. Deep residual learning for image recognition. in *Proceedings of the IEEE conference on computer vision and pattern recognition* 770–778 (2016).

3. Cho, K., Merrienboer, B. van, Bahdanau, D. & Bengio, Y. On the Properties of Neural Machine Translation: Encoder-Decoder Approaches. Preprint at https://doi.org/10.48550/arXiv.1409.1259 (2014).

4. Papale, P., Wang, F., Self, M. W. & Roelfsema, P. R. An extensive dataset of spiking activity to reveal the syntax of the ventral stream. *Neuron* **0**, (2025).